\documentclass[a4paper]{article}

\usepackage{INTERSPEECH2019}
\usepackage{amsmath}
\usepackage{mathtools,xparse}
\usepackage{booktabs}
\usepackage{graphics}
\usepackage{multirow}
\DeclarePairedDelimiter\abs{\lvert}{\rvert}%
\title{3-D Feature and Acoustic Modeling for Far-Field Speech Recognition}
\name{Anurenjan Purushothaman, Anirudh Sreeram, Sriram Ganapathy\thanks{This work was partly funded by the project grants from Samsung Research India, Bangalore.}}
\address{
 Learning and Extraction of Acoustic Patterns (LEAP) lab, Indian Institute of Science, Bangalore.}
\email{\{anurenjanr, sanirudh, sriramg\}@iisc.ac.in}

\begin{document}

\maketitle
\begin{abstract}
Automatic speech recognition in multi-channel reverberant conditions is a challenging task. The conventional way of suppressing the reverberation artifacts involves a beamforming based enhancement of the multi-channel speech signal, which is used to extract spectrogram based features for a neural network acoustic model.  In this paper, we propose to extract features directly from the multi-channel speech signal using a multi variate autoregressive (MAR) modeling approach, where the correlations among all the three dimensions of time, frequency and channel are exploited. The MAR features are fed to a convolutional neural network (CNN) architecture which performs the joint acoustic modeling on the three dimensions. The 3-D CNN architecture allows the combination of multi-channel features that optimize the speech recognition cost compared to the traditional beamforming models that focus on the enhancement task. Experiments are conducted on the CHiME-3 and REVERB Challenge dataset using multi-channel reverberant speech. In these experiments, the proposed 3-D feature and acoustic modeling approach provides significant improvements over an ASR system trained with beamformed audio (average relative improvements of $10$\% and $9$\%  in word error rates for CHiME-3 and REVERB Challenge datasets respectively). 
\end{abstract}
\vspace{0.3cm}

\noindent\textbf{Index Terms}: MAR Modeling, Riesz Envelopes, Multi-Channel Features, Beamforming, 3-D CNN models.
\vspace{-0.0cm}

\section{Introduction}
Automatic speech recognition (ASR) systems find widespread use in applications like human-machine interface, virtual assistants, smart speakers etc, where the input speech is often reverberant and noisy. The ASR performance has improved dramatically over the last decade with the help of deep learning models \cite{yu2016automatic}. However, the degradation of the systems in presence of noise and reverberation continues to be a challenging problem due to the low signal to noise ratio \cite{hain2012transcribing}.  For \textit{e.g.} Peddinti \textit{et al,} \cite{peddinti2017low} reports a $75\%$ rel. increase in word error rate (WER) when signals from a far-field array microphone are used in place of those from headset microphones in the ASR systems, both during training and testing. This degradation could be primarily attributed to reverberation artifacts \cite{yoshioka2012making, kinoshita2013reverb}. The availability of multi-channel signals can be leveraged for alleviating these issues as most of the real life far-field speech recordings are captured by a microphone array.

Previously, many works have focused on far-field speech recognition using multiple microphones \cite{far1,far2,far3,far4}. The  traditional approach to multi channel far-field ASR combines all the available channels by beamforming \cite{anguera2007acoustic} and then processes the resulting single channel signal effectively. The technique of beamforming attempts to find the time delay between channels and boosts the signal by weighted and delayed summation of the individual channels \cite{wolfel2009distant,delcroix2015strategies}. This approach is still the most successful system for ASR in multi-channel reverberant environments \cite{barker2018fifth}.  

In this paper, we propose an approach to avoid the beamforming step by directly processing the multi-channel features within the ASR framework.  A feature extraction step is proposed that is based on multi-variate autoregressive (MAR) modeling  exploiting the joint correlation among the three dimensions of time, frequency and channel present in the signal. A novel neural network architecture for multi-channel ASR is also proposed that contains network-in-network (NIN) in a 3-D convolutional neural network (CNN) architecture. With several ASR experiments conducted on CHiME-3 \cite{chime3} and REVERB challenge dataset \cite{rev1,rev2}, we show that the proposed approach to multi-channel feature and acoustic modeling improves significantly  over a baseline system using conventional beamformed audio with mel filter bank energy features.

The rest of the paper is organized as follows. The related prior works are discussed in Section~\ref{sec:prior_work}. The details about the proposed 3-D features are provided in Section~\ref{sec:mar_feat}. Section~\ref{sec:neural_network} elaborates the proposed model architecture for multi-channel ASR. The ASR experiments and results are reported in Section~\ref{sec:expt}, which is followed by a  summary in Section~\ref{sec:summary}.   
\vspace{-0.0cm}

\section{Retaled Prior Work}\label{sec:prior_work}
While the original goal of beamforming \cite{anguera2007acoustic} is directed towards signal enhancement, the beamforming cost can be modified for maximizing the likelihood \cite{seltzer2004likelihood}. With the advent of neural network based acoustic models, multi-channel acoustic models have also been explored. Recently, Swietojanski \textit{et al} \cite{swietojanski2014convolutional} proposed the use of features from each channel of the multi-channel speech directly as input to a convolutional neural network based acoustic model. Here, the neural network is seen as a replacement for conventional beamformer. Joint training of a more explicit beamformer with the neural network acoustic model has been proposed by Xiao \textit{et al.,} \cite{xiao2016deep}. Training of neural networks, which operate on the raw signals that are optimized for the discriminative cost function of the acoustic model, has also been recently explored. These approaches are termed as \textit{Neural Beamforming} approaches as the neural network acoustic model subsumes the functionality of the beamformer \cite{sainath2017multichannel, ochiai2017unified}.

Previously, we had explored the use of 3-D CNN models  in  \cite{ganapathy}, where the network was fed with the spectrogram features of all channels.  Separately, a multi-band feature extraction using autoregressive modeling was proposed for deriving noise robust features from single channel speech \cite{marsri,ganapathy2018far}. 

In this paper, we use the multi-variate autoregressive modeling features (MAR) from the microphone array to derive 3-D features. We also extend the previous work on  3-D CNN models \cite{ganapathy}  with a newer architecture that combines the multiple channel features in a NIN framework. 
\begin{figure*}
  \centering
  \includegraphics[width=\textwidth,height=5.7cm]{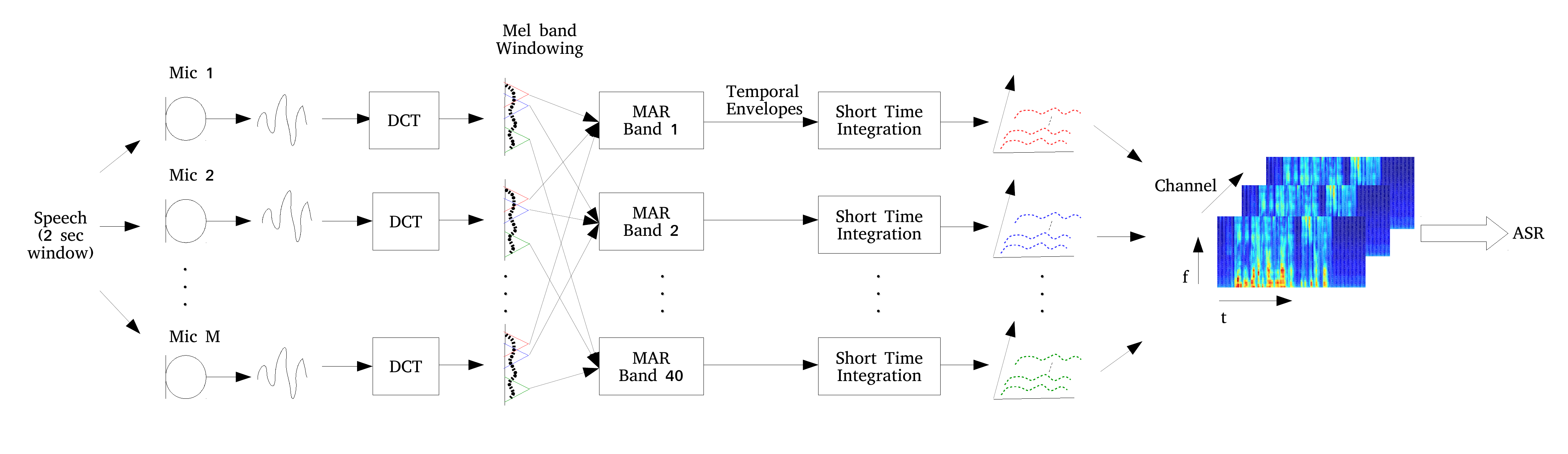}
      \caption{Block schematic of the 3-D feature extraction method using MAR modeling.}
  
  \label{fig2}
  \vspace{-0.3cm}
\end{figure*}
\section{3-D MAR features}\label{sec:mar_feat}
Multi variate autoregressive (MAR) modeling  was proposed to derive robust features in the joint time-frequency domain \cite{marsri}. In this case, the model assumes that the discrete cosine transform (DCT) components of different frequency sub-bands can be jointly expressed in a vector linear prediction process.  The frame work relies on frequency domain linear prediction which states that linear prediction applied on frequency domain estimates the envelopes of the signal \cite{athineos2003frequency,sriphd}.  We review the mathematical model of the MAR feature extraction and go on to propose the model for multi-channel feature extraction. 
\subsection{Discrete Cosine Transform (DCT)}
Let $x[n]$ denote a discrete sequence. The DCT $[y[k]$ is given by,
\begin{equation}
        y[k] = a[k]\sum_{n=0}^{N-1}x[n]cos\left(\frac{\left(2n +1\right)\pi k}{2N}\right), k = 0, 1,\hdots, N-1
\end{equation}
where $a[k]=1$ for $k=0$, $\sqrt{2}$ otherwise.
\subsection{Frequency Domain Linear Prediction}
FDLP is the frequency domain dual of Time Domain Linear Prediction (TDLP). Just as TDLP estimates the spectral envelope of a signal, FDLP estimates the temporal envelope of the signal, i.e. square of its Hilbert envelope \cite{analytic}. Temporal envelope is given by the inverse Fourier transform of the autocorrelation function of DCT.
\begin{equation}
    e(t) = F^{-1}\left \{Autocorr(y[k])\right \}
\end{equation}
We use the autocorrelation of the DCT coefficient to predict the temporal envelope of the signal. One of the inherent property of linear prediction is that, it tries to approximate the peaks very well. The FDLP model  tries to preserve the peaks in temporal domain. 
\subsection{Multi-channel feature extraction}
For each channel, we use a $2$ second window for DCT computation. We partition the DCT signal in frequency domain by multiplying with a window $w_i[k]$. The window functions  $w_i[k]$ are chosen to be uniformly spaced in the mel-scale and have a Gaussian shape \cite{o1987speech}. Let $y_i[k]$ denotes the $i^{th}$ sub band DCT. The corresponding sub bands of all the channels are appended to form a vector $\textbf{y}_i^k$.
\begin{equation}
    \textbf{y}_i^k={\begin{bmatrix}
    y_i^1[k]~y_i^2[k]~ \hdots~y_i^C[k]
\end{bmatrix}}^T
\end{equation}
where $y_i^1[k]$ denotes the windowed $i^{th}$ sub band from the first channel and $C$ is the  number of available channels. We perform a vector linear prediction on the signal $\textbf{y}_i^k$. This will reveal the multivariate autoregressive model of the signal.  
\subsection{Multi variate Autoregressive Modeling}
The $C$ dimensional wide  sense stationary vector process $\mathbf{y}_i^k$ is said to be autoregressive \cite{vaidyanathan2007theory} if it is generated by a recursive difference equation of the form
\begin{equation}
   \mathbf{y}_i^k=-\sum_{l=1}^N\mathbf{D}_l\mathbf{y}_i^{k-l} + \bm{\epsilon}^k
\end{equation}
where $\bm{\epsilon}^k$ is an $C$ dimensional white noise random process with a covariance matrix $\mathbf{\Sigma_{\epsilon}}$ and the MAR coefficients $\mathbf{D}_l$ are square matrices of size $C$ which characterize the model \cite{lutp}.

We use the autocorrelation method for the solution of normal equation to find the model parameters \cite{vaidyanathan2007theory}. The forward prediction polynomial is given by
\begin{equation}
    \mathbf{H}[z]=\mathbf{I}_L + \mathbf{D}_1z^{-1} + \mathbf{D}_2z^{-2} + ... + \mathbf{D}_Nz^{-N}
\end{equation}
where $z$ represents complex time domain variable \cite{kumaresan1999model}. The optimal predictor is solved by minimizing the mean square error as follows.
\begin{equation}
\footnotesize{
\begin{bmatrix}
    \mathbf{R}(0) & \ \dots & \mathbf{R}(N-1) \\
    \mathbf{R}(-1) & \ \dots & \mathbf{R}(N-2) \\
    \vdots & \  \ddots & \vdots \\
    \mathbf{R}(-N+1) & \ \dots & \mathbf{R}(0)
\end{bmatrix}
\begin{bmatrix}
    \mathbf{D}_1 \\
    \mathbf{D}_2 \\
    \vdots  \\
    \mathbf{D}_N
\end{bmatrix}
=-
\begin{bmatrix}
    \mathbf{R}(-1) \\
    \mathbf{R}(-2) \\
    \vdots  \\
    \mathbf{R}(-N)
\end{bmatrix}}
\end{equation}
where $\mathbf{R}(q)$ is the autocorrelation matrix of the WSS process $\mathbf{y}_i^k$ for lag $q$ given by
\begin{equation}
    \mathbf{R}(q)=E[\mathbf{y}_i^k{\mathbf{y}_i^{k-q}}^{~T}]
\end{equation}
Here, $E$ denotes the expectation operator and ${\mathbf{y}_i^{k-q}}^{~T}$ represents the transpose of $\mathbf{y}_i^{k-q}$. The estimate of the error covariance matrix $\mathbf{\hat{\Sigma}}_e$ which is Hermitian is given by 
\begin{equation}
    \mathbf{\hat{\Sigma}}_e=\mathbf{R}(0) + \mathbf{R}(1)\mathbf{D}_1 + \mathbf{R}(2)\mathbf{D}_2 + ... + \mathbf{R}(N)\mathbf{D}_N
\end{equation}
\begin{figure*}[t!]
  \centering
  \includegraphics[width=\textwidth,height=4.0cm]{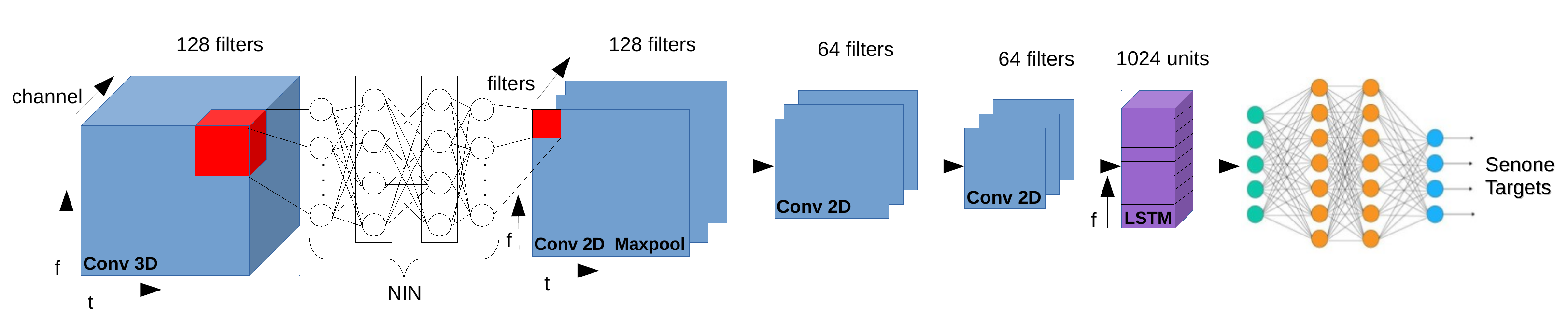}
  \caption{3-D Conv-LSTM architecture used in multi-channel ASR which has NIN 1\textsuperscript{st} layer performing 3-D CNN, 2-D CNN, and LSTM layers.}
  \label{fig1}
  \vspace{-0.3cm}
\end{figure*}
\subsection{Envelope Estimation}
The goal of performing linear prediction in our case is to estimate the temporal envelopes. In this paper, the input $\mathbf{y}_i^k$ denotes DCT coefficients indexed by $k$ for the sub band $i$ from all channels. The corresponding Hilbert envelopes are estimated using MAR modeling. If $\mathbf{e}_{i}[n]$ denotes the multi dimensional Riesz envelope (extension of Hilbert envelope to 2-D signals) \cite{riesz} of multi channel speech for sub-band $i$ $(\mathbf{e}_{i}[n] = {\begin{bmatrix}
    e_i^1[n]~e_i^2[n]\dots e_i^C[n]\end{bmatrix}}^T)$, then the MAR estimate of the Riesz envelope is given by the following equation 
\begin{equation}
    \hat{\mathbf{e}}_{i}[n] = diag(\mathbf{H}[n]^{-1}\hat{\mathbf{\Sigma}}_e\mathbf{\Bar{H}}[n]^{-1})
\end{equation}
where $\mathbf{H}[n] = \mathbf{H}[z]|_{z=\exp^{-j2 \pi n}}$ with $\mathbf{H}[z]$ given by equation (5) and $\hat{\mathbf{e}}_{i}[n] = {\begin{bmatrix}
    \hat{e}_i^1[n]~\hat{e}_i^2[n]\dots \hat{e}_i^C[n]\end{bmatrix}}^T$ . By estimating $\hat{\mathbf{e}}_{i}[n]$ for each sub band, we reconstruct the temporal envelopes of all the channels and all sub bands. Re-arranging sub band envelopes gives the 3-D feature representation.
\subsection{Gain normalization}
In order to reduce the dynamic range of envelopes we normalize the magnitude of envelope over the two second computation window. This has the effect of suppressing additive noise artifacts \cite{sriphd}. It is to be noted that the gain normalization of band energies is done for CHiME-3 dataset (which has additive noise), but not on REVERB Challenge dataset.
\subsection{Multi channel Feature Extraction for ASR using MAR}
The block schematic of the proposed multi channel feature extraction is shown in Figure 1. Long segments of speech from each channel  are taken (non- overlapping 2 sec duration) and are transformed by DCT. The full band DCT is windowed into overlapping 40 sub bands. This data is fed into the MAR feature extraction block. The estimation procedure of the multivariate AR model is applied and model parameters are estimated. We chose $N = 107$. The sub band MAR envelopes $\hat{\textbf{e}}_i[n]$ are integrated with a Hamming window over a 25 ms window with a 10 ms shift. The integration in time of the sub band envelopes yields an estimate of the MAR spectrogram of the input speech signal.
\vspace{-0.2cm}
\section{Model Architectures}\label{sec:neural_network}
The proposed 3-D CLSTM architecture is shown in Figure 2. The input data consists of 21 frames of 40 bands from all the  $C$ channels. The input data to 3-D CLSTM model is a 3-D tensor of size $C\times21\times40$ in the first layer, followed by a 2-D CNN layer with 128 kernels of size 1$\times$3$\times$3 in the second. This is followed by maxpooling and two 2-D CNN layers with 64 filters of kernel size 1$\times$3$\times$3. The output of the convolution layers is fed to an LSTM \cite{lstm} which performs frequency recurrence. This is followed by a fully connected layer \cite{mlp}, which predicts the senone classes. Dropout \cite{dropout} and batch normalization \cite{batch} are used for regularization. The model training is performed using Pytorch software \cite{pytorch}.

In order to enhance the learning of the non linearity in the filters of the 3-D CNN layer, we use the Network in Network (NIN) \cite{nin} architecture. The NIN is used in the first layer to learn the non linearity present in the filters. The first layer performs the equivalent of neural beamforming while successive layers have only 2-D $t \times f$ representation. 

The 2-D CLSTM architecture used in the case of  beamformed audio, is a special case of the proposed 3-D CLSTM architecture, where the input is a 2-D spectrogram of size 21$\times$40 and a normal 2-D Convolution is performed in the initial layer. The rest of the network architecture starting from layer-2 of model (shown in Figure 2) is used for the 2-D CLSTM model on beamformed audio features.
\begin{center}
\begin{table}[t!]
\caption{Word Error Rate (\%) in CHiME-3 dataset for beamformed FBANK features using different model architectures.}
\vspace{-0.2cm}
\centering
\begin{tabular}{@{}l|cc|cc@{}}
\toprule
\multicolumn{1}{c|}{\multirow{2}{*}{Experiments}} & \multicolumn{2}{c|}{Dev.}                                                       & \multicolumn{2}{c}{Eval.}                                                     \\ \cmidrule(l){2-5} 
\multicolumn{1}{c|}{}   & \multicolumn{1}{l}{Real.} & \multicolumn{1}{l|}{Simu.} & \multicolumn{1}{l}{Real.} & \multicolumn{1}{l}{Simu.}  \\ \midrule
DNN                                         &11.6                     &13.4                                            &20.5                      &20.9                                          \\
CNN2D                                       & 9.4                  &12.3                                       &17.7                     &19.1                                         \\
$~$ + Dropout                                       &8.6                    &11.4                                         &16.3                     &17.7                                       \\
$~$ + Batchnorm, Adam                                          &{8.4}                      &11.3                                           &16.0                    &17.7                                      \\
$~$ + LSTM                                        &8.4                      &11.0                             &16.0                     &17.1                               \\

\bottomrule
\end{tabular}

\label{table:1}
\vspace{-0.4cm}
\end{table}
\end{center}
\vspace{-1cm}
\section{Experiments and Results}\label{sec:expt}
The experiments are performed on CHiME-3 and REVERB Challenge datasets. For the baseline model, multiple architectures are experimented using beamformed FBANK (40 band mel spectrogram with frequency range from 200 Hz to 6500 Hz) as the features (Table 1). The 2-D CNN architecture gives a significant improvement over the DNN. Adding dropouts helped improve the performance further. Batch normalization and Adam optimizer also showed marginal improvement over the 2-D CNN model with dropout. Finally, we propose a new CLSTM architecture with the LSTM recurring over frequency. This served as the baseline for our experiments on the multi-channel data.

We also perform experiments with multi-band feature extraction \cite{marsri} on the beamformed audio (BF-MB), using the 2-D CLSTM architecture.
\subsection{CHiME-3 ASR}
The CHiME-3 dataset \cite{chime3} for the ASR has multiple microphone tablet device recording in four different environments, namely, public transport (BUS), cafe (CAF), street junction (STR) and pedestrian area (PED). For each of the above  environments real and simulated data are present. The real data consists of $6$ channel recordings from WSJ0 corpus sampled at $16$ kHz spoken in the four varied environments. The simulated data was constructed by mixing clean utterances with the environment noise. The training dataset consists of $1600$ (real) noisy and $7138$ (simulated) noisy utterances from $83$ speakers. The development (Dev) and evaluation (Eval) datasets consists of $1640~(410\times4)$ from $4$ speakers and $1320~(330\times4)$ from $4$ other speakers real Dev and Eval data respectively. Identically sized simulated Dev and Eval datasets are also present.

\begin{table}[t!]
\caption{Word Error Rate (\%) in CHiME-3 dataset for different CNN configurations using 3-D MAR features.}
\vspace{-0.2cm}
\resizebox{\columnwidth}{!}{%
\begin{tabular}{@{}l|ccc|ccc@{}}
\toprule
\multicolumn{1}{c|}{\multirow{2}{*}{3-D CNN Config.}}                        & \multicolumn{3}{c|}{Dev.} & \multicolumn{3}{c}{Eval.} \\ \cmidrule(l){2-7} 
\multicolumn{1}{c|}{}                                                    & Real.    & Simu.   & Avg.   & Real.    & Simu.   & Avg.   \\ \midrule
3D kernels (2 layers)                                                     & 9.9     & 9.8    & 9.9    & 19.2    & 12.7   & 15.9   \\
3-D kernels (1 layer)                                                       & 10.1    & 10.5   & 10.3   & 19.2    & 14.0   & 16.6   \\
+ NIN (1 hidden layer)                                                   & 10.2    & 10.3   & 10.3   & 19.8    & 13.6   & 16.7   \\
\begin{tabular}[c]{@{}l@{}}+ NIN (2 hidden layer),\\ Dropout\end{tabular} & 9.3     & 9.4    & \textbf{9.3}    & 17.3    & 12.7   & \textbf{15.0}   \\ \bottomrule
\end{tabular}
}
\label{table:1}
\vspace{-0.2cm}
\end{table}

The effect of different CNN configurations in the first two layers of the proposed 3-D CLSTM architecture is reported in Table 2. Although removing the channel level information in the first layer (L1) reduces the performance of the ASR compared to removing it in the first two layers (L1+L2), with NIN and dropout the former becomes better. Performance of the ASR improves over the baseline (BF-FBANK) by using BF-MB features as shown in Table 3.

We compare the performance of the proposed 3-D Feature and acoustic model, named as MC-MAR with the baseline (Table 3). In the multi-channel experiments, 5 channel recordings are taken and multi-channel features are extracted. All the filter bank and MAR features are extracted by keeping the number of bands as 40 and the data is trained using the proposed 3-D CLSTM architecture.

\begin{table}[t!]
\caption{Word Error Rate (\%) in CHiME-3 dataset for different feature and configuration.}
\vspace{-0.2cm}

\resizebox{\columnwidth}{!}{%
\begin{tabular}{@{}c|ccc|ccc@{}}
\toprule
Experiments                                                      & \multicolumn{3}{c|}{Dev.}     & \multicolumn{3}{c}{Eval.}    \\ \midrule
\begin{tabular}[c]{@{}c@{}}Feature\\ (Architecture)\end{tabular} & Real. & Simu. & Avg.         & Real. & Simu. & Avg.          \\ \midrule
\begin{tabular}[c]{@{}c@{}}BF-FBANK\\ (2-D CLSTM)\end{tabular}   & 8.4   & 11.0  & 9.7          & 16.0  & 17.1  & 16.6          \\
\begin{tabular}[c]{@{}c@{}}BF-MB\\ (2-D CLSTM)\end{tabular}     & 8.4   & 10.9  & 9.6          & 15.1  & 16.5  & 15.8          \\
\begin{tabular}[c]{@{}c@{}}MC-FBANK\\ (3-D CLSTM)\end{tabular}   & 9.7   & 9.9   & 9.8          & 19.7  & 13.0  & 16.3          \\
\begin{tabular}[c]{@{}c@{}}MC-MAR\\ (3-D CLSTM)\end{tabular}     & 9.3   & 9.4   & \textbf{9.3} & 17.3  & 12.7  & \textbf{15.0} \\ \bottomrule
\end{tabular}
}
\label{table:1}
\vspace{-0.6cm}
\end{table}

The results for multi-channel ASR experiments on CHiME-3 dataset are shown in Table 1, 2, 3 \& 4. It can be seen that the proposed 3-D features and 3-D CLSTM model has average relative improvement of $10$ \% in WER for CHiME-3 dataset. The 3-D CNN model on FBANK features (MC-FBANK) \cite{ganapathy} also shows a marginal improvement over the beamformed FBANK (BF-FBANK) features.

\begin{table}[t!]
\caption{Word Error Rate (\%) for different noise conditions in CHiME-3 dataset on MC-MAR with 3-D CLSTM architecture (Baseline with 2-D CNN architecture shown in paranthesis.)}
\vspace{-0.2cm}

\centering

\begin{tabular}{@{}cl|cc|c|c@{}}
\toprule
\multicolumn{2}{c|}{\multirow{2}{*}{Cond.}} & \multicolumn{2}{c|}{Dev.}         & \multicolumn{2}{c}{Eval.} \\ \cmidrule(l){3-6} 
\multicolumn{2}{c|}{}                       & \multicolumn{1}{c|}{Real.} & Simu. & Real.         & Simu.        \\ \midrule
\multicolumn{2}{c|}{BUS}                     & \multicolumn{1}{c|}{12.0 (9.9)} & 8.1 (9.1)  & 23.5 (20.3)         & 9.3 (11.6)         \\
\multicolumn{2}{c|}{CAF}                     & \multicolumn{1}{c|}{8.7 (8.4)}  & 11.9 (14.0) & 16.0 (16.0)         & 13.9 (19.0)        \\
\multicolumn{2}{c|}{PED}                     & \multicolumn{1}{c|}{7.1 (6.8)}  & 8.0 (9.4)  & 18.0 (15.6)         & 13.2 (18.1)        \\
\multicolumn{2}{c|}{STR}                     & \multicolumn{1}{c|}{9.3 (8.6)}  & 9.5 (11.6)  & 11.8 (12.1)         & 14.3 (19.8)        \\ \bottomrule
\end{tabular}

\label{table:2}
\vspace{-0.2cm}
\end{table}
\vspace{-0.3cm}
\subsection{REVERB Challenge ASR}
The REVERB Challenge dataset \cite{rev3} for ASR consists of 8 channel recordings. Real and Simulated noisy speech data are present. Simulated data is comprised of reverberant utterances generated based on the WSJCAM0 corpus \cite{rev1}. These utterances were artificially distorted by convolving clean WSJCAM0 signals with measured room impulse responses (RIRs) and adding noise at an SNR of 20 dB. SimData simulated six different reverberation condition. Real data, which is comprised of utterances from the MC-WSJ-AV corpus \cite{rev2} consists of utterances spoken by human speakers in a noisy and reverberant room. The training set consists of 7861 uttrances (92 speakers) from the clean WSJCAM0 training data by convolving the clean utterances with 24 measured RIRs and adding noise at an SNR of 20 dB. The development (Dev) and evaluation (Eval) datasets consists of 1663 (1484 simulated and 179 real) and 2548 (2176 simulated and 372 real) utterances respectively. The Dev and Eval datasets have 20 and 28 speakers respectively.

\begin{table}[t!]
\caption{Word Error Rate (\%) in REVERB dataset for various feature extraction methods.}
\vspace{-0.2cm}

\resizebox{\columnwidth}{!}{%
\begin{tabular}{@{}c|ccc|ccc@{}}
\toprule
Experiments                                                      & \multicolumn{3}{c|}{Dev.}     & \multicolumn{3}{c}{Eval.}    \\ \midrule
\begin{tabular}[c]{@{}c@{}}Feature\\ (Architecture)\end{tabular} & Real. & Simu. & Avg.          & Real. & Simu. & Avg.          \\ \midrule
\begin{tabular}[c]{@{}c@{}}BF-FBANK\\ (2-D CLSTM)\end{tabular}   & 21.8  & 7.1   & 14.4          & 23.8  & 7.6   & 15.7          \\
\begin{tabular}[c]{@{}c@{}}BF-MB\\ (2-D CLSTM)\end{tabular}     & 20.7  & 7.3   & 14.0          & 21.6  & 7.6   & 14.6          \\
\begin{tabular}[c]{@{}c@{}}MC-FBANK\\ (3-D CLSTM)\end{tabular}   & 21.5  & 7.9   & 14.7          & 23.4  & 8.1   & 15.7          \\
\begin{tabular}[c]{@{}c@{}}MC-MAR\\ (3-D CLSTM)\end{tabular}     & 19.6  & 7.7   & \textbf{13.6} & 20.5  & 8.0   & \textbf{14.3} \\ \bottomrule
\end{tabular}
}
\label{table:2}
\vspace{-0.6cm}
\end{table}

The results for multi-channel ASR experiments on REVERB Challenge are shown in Table 5. It can be seen that the proposed 3-D features and 3-D CLSTM model provides average relative improvement of $9$ \% in WER for REVERB Challenge dataset over the BF-FBANK 2-D CLSTM baseline. The trends observed in REVERB Challenge are also similar to those seen in CHiME-3 dataset.

\vspace{-0.25cm}

\section{Summary}\label{sec:summary}
In this paper, we propose a new framework of multi-channel features using MAR modeling in the frequency domain. We also propose 3-D CNN model for neural beamforming. Various speech recognition experiments were performed on the CHiME-3 dataset as well as the REVERB Challenge dataset. The main conclusion of our experiments is that using multi-channel acoustic model, the performance of ASR can be improved for far-field speech. The analysis also highlights the incremental benefits achieved for various feature and model architecture combinations.
\bibliographystyle{IEEEtran}
\bibliography{mybib}
\end{document}